\documentclass[journal,10pt,twocolumn,twoside]{IEEEtran}
\usepackage{algorithm,algorithmic}
\usepackage{amsmath,graphicx,amsthm}
\usepackage{epstopdf}
\usepackage{amssymb}
\usepackage{multirow}
\usepackage{epstopdf}
\usepackage{graphicx}
\usepackage{xcolor}
\usepackage{booktabs}

\newcommand{\bs}{\boldsymbol}
\newcommand{\mb}{\mathbf}
\DeclareMathOperator*{\argmin}{arg\,min}

\def\Expect{\mathbb{E}}
\def\R{\mathbb{R}}
\def\N{\mathcal{N}}
\def\lone{\ell_1}

\def\U{\mathcal{U}}

\newtheorem{Theorem}{Theorem}

\newtheorem{Finding}{Finding}
\begin{document}
%
\title{Near optimal compressed sensing without priors: Parametric SURE Approximate Message Passing}
%
%
%
\author{Chunli Guo\authorrefmark{1},~\IEEEmembership{Student Member,~IEEE,}
        and Mike E. Davies,~\IEEEmembership{Senior Member,~IEEE}\\
        
\thanks{The authors are with the Institute for Digital Communication and with the Joint Research Institute for Signal and Image
Processing, Edinburgh University, King’s Buildings, Mayfield Road, Edinburgh EH9 3JL, UK (e-mail: c.guo@ed.ac.uk, mike.davies@ed.ac.uk).

C.Guo receives her PhD Scholarship from the Maxwell Advanced Technology Fund at the University of Edinburgh. MED acknowledges support of his position from the Scottish Funding Council and their support of the Joint Research Institute with the Heriot-Watt University as a component part of the Edinburgh Research Partnership in Engineering and Mathematics. 

Some of the work presented here will be presented at the European Signal Processing Conference, Lisbon, Portugal, September 2014 \cite{guo2014}. }}


\maketitle

\begin{abstract}
Both theoretical analysis and empirical evidence confirm that the approximate message passing (AMP) algorithm can be interpreted as recursively solving a signal denoising problem: at each AMP iteration, one observes a Gaussian noise perturbed original signal. Retrieving the signal amounts to a successive noise cancellation until the noise variance decreases to a satisfactory level. In this paper we incorporate the Stein's unbiased risk estimate (SURE) based parametric denoiser with the AMP framework and propose the novel parametric SURE-AMP algorithm. At each parametric SURE-AMP iteration, the denoiser is adaptively optimized within the parametric class by minimizing SURE, which depends purely on the noisy observation. In this manner, the parametric SURE-AMP is guaranteed with the best-in-class recovery and convergence rate. If the parameter family includes the families of the mimimum mean squared error (MMSE) estimators, we are able to achieve the Bayesian optimal AMP performance without knowing the signal prior. In the paper, we resort to the linear parameterization of the SURE based denoiser and propose three different kernel families as the base functions. Numerical simulations with the Bernoulli-Gaussian, $k$-dense and Student's-t signals demonstrate that the parametric SURE-AMP does not only achieve the state-of-the-art recovery but also runs more than 20 times faster than the EM-GM-GAMP algorithm.

\end{abstract}


\IEEEpeerreviewmaketitle

\section{Introduction}
\IEEEPARstart{C}{ompressed} sensing (CS) refers to a technique that retrieves the information of a sparse or compressible signal with a sampling ratio far below the Nyquist rate. Given the measurement matrix $\bs\Phi\in\R^{m\times n}$, $m \ll n$ and the noisy observation $\mb y=\bs\Phi\mb x_o + \mb w \in\R^{m}$, where $\mb w$ is the noise vector with i.i.d Gaussian random entries $w_i\in\N(w_i; 0, \sigma^2_w)$, the CS reconstruction task is to solve the severely under-determined linear system to recover the original signal $\mb x_o\in\R^{n}$. Over the past decade, CS techniques have been found to be valuable in a wide range of practical scenarios, including the natural image processing, the medical imaging, the radar tasks and the astro-imaging, to name just a few. Since signals encountered in practice are normally very large, developing efficient reconstruction algorithm with low computational cost is one of the most discussed topics in the CS community.

Within all existing CS reconstruction algorithms, the approximate message passing (AMP) algorithm and its variants exhibit the attractive reconstruction power and low computational complexity. First introduced by Donoho et. al in \cite{AMP}, AMP based algorithms generally take a simple iterative form:
\begin{align}
\label{eq:AMP}
\mb r^t &=  \hat{\mb x}^t + \bs\Phi^T\mb z^t\\ \label{eq:AMP2}
\hat{\mb x}^{t+1} &=  \eta_{t}(\mb r^t)\\ \label{eq:AMP3}
\mb z^{t+1}& =  \mb y -\bs\Phi\hat{\mb x}^{t+1} +\frac{1}{\gamma}\mb z^t <\eta'_t(\mb r^t)>
\end{align}
where $\gamma=m/n$ is the sampling ratio. Initialized with $\hat{\mb x}^0 =\mb 0$ and $\mb z^0 = \mb y$, AMP iteratively produces an estimation of the original signal $\hat{\mb x}^t$ with a scalar non-linear function $\eta_t(\cdot)$, which is applied elementwise to $\mb r^t$. With $\bs\Phi$ being the Gaussian random measurement matrix, $\mb r^t$ at each AMP iteration can be effectively modelled as the original signal with some Gaussian perturbation in the large system limit. To be specific, we approximately have $\mb r^t \approx \mb x_o + \sqrt{\tau_t} \mb z^t$, $z_i\in\N(z_i; 0,1)$, where $\tau_t$ is the effective noise variance \cite{AMP,DonohoanalysisAMP}. Then the non-linearity $\eta_t(\cdot)$ essentially acts as a denoising function to remove the Gaussian noise $\sqrt{\tau_t}\mb z^t$.

In the original AMP paper \cite{AMP,Donoho10messagepassing}, the denoising is achieved with the simple soft thresholding function $\eta(r,c)=sign(r)(\lvert r\rvert -c)_{+}$, where $(r)_{+} = r\mathbb{I}(r\geq 0)$. $\mathbb{I}(\cdot)$ is the indicator function. The corresponding AMP algorithm, dubbed as the $\lone$-AMP, is proved to have the identical phase transition for sparse signal reconstruction as the $\lone$-minimization approach \cite{DonohoanalysisAMP}. Despite the fact that the noisy vector $\mb r^t$ has multiple i.i.d distributed elements, the $\lone$-AMP treats the denoising as a 1-d problem and utilizes the element-wise soft-thresholding function as the denoiser. However, since the true signal pdf is visible in the noisy estimate in the large system limit and the effective noise variance is estimated at each AMP iteration, we should be able to exploit such information to achieve better recovery than the $\lone$-AMP. 

It is well known that when $p(\mb x_o)$ is known, the optimal denoising with the least mean square error (MSE) is achieved by applying the minimum mean squared error (MMSE) estimator. Consequently, the AMP algorithm which deploys the MMSE estimator for denoising achieves the best reconstruction in the least square sense, and is denoted as the Bayesian optimal AMP (BAMP) algorithm \cite{BayAMP}. However, the requirement of $p(\mb x_o)$ to be known in advance can be restrictive in practice. The advantages and limitation of BAMP also motive us to find an alternative approach which is able to fill the gap between the $\lone$-AMP and the BAMP, or even performs as well as BAMP without knowing the signal distribution a priori.   



\subsection{Main contributions}
In the large system limit, the true prior for $\mb x_o$ at each AMP iteration is essentially embedded in the data $\mb r^t$, which is the convolution of the original signal with the Gaussian noise kernel. To improve the recovery, we could either estimate the pdf and then deduce the associated MMSE estimator, or directly optimize the denoising. In this paper, we adopt the latter approach and propose the parametric SURE-AMP algorithm. Realizing the recursive denoising nature of the AMP iteration, we introduce a class of parameterized denoising functions to the generic AMP framework. At each iteration, the denoiser with the least MSE is selected within the class by optimizing the free parameters. In this manner, the parametric SURE-AMP algorithm adaptively chooses the best-in-class denoiser and achieves the best possible denoising within the parametric family at each iteration. When the denoiser class contains all possible MMSE estimators for a specific signal, the parametric SURE-AMP is expected to achieve the BAMP recovery without knowing the signal prior.  


The key feature of the parametric SURE-AMP algorithm is that the denoiser optimization does not require prior knowledge of $p(\mb x_o)$. To make this possible, we resort to the Stein's unbiased risk estimate (SURE) based parametric least squarer denoiser construction. There exists a rich literature on signal denoising with SURE ~\cite{Donoho1995,Pesquet1997, Benazza2005, Luisier2006, Unser2007,Blu2007,raphan2011least}. Since SURE is the unbiased estimate of MSE, the pursuit of the best denoiser with the least MSE is nothing more than minimizing the corresponding SURE. More importantly, for Gaussian noise corrupted signal, the calculation of SURE depends purely on the sampled average of the noisy data \cite{stein1981}. By leveraging the large system limit, the best-in-class denoiser can be determined without the prior information \cite{raphan2011least}. 

The success of the parametric SURE-AMP relies heavily on the parameterization of the denoiser class. The number of parameters as well as the linearity determine the optimization complexity. In this paper, we restrict ourselves to the linear combination of non-linear kernel functions as the denoiser structure. The non-linear parameters of the kernel functions are set to have a fixed ratio with the effective noise variance. The linear weights for the kernels are optimized by solving a linear system of equations. We presented two types of piecewise linear kernel family and one exponential kernel family for both sparse and heavy-tailed signal reconstruction. The numerical simulation with the Bernoulli-Gaussian (BG), $k$-dense and Student's-t signals show that with a limited number of kernel functions, we are able to adaptively capture the evolving shape of the MMSE estimator and achieves the state-of-art performance in the sense of reconstruction quality and computational complexity.

\subsection{Related literature}
The pre-requisite of the signal prior to implement BAMP has been noticed by several research groups. To tackle this limitation, the prior estimation step was proposed to be incorporated within the AMP framework. In \cite{Vila2011, Vila2013, MagicMatrix, MagicMatrixExtend}, a Gaussian mixture (GM) model is used as the parametric representation of $p(\mb x_o)$. The expectation-maximization (EM) approach is deployed to jointly learn the prior along with recovering $\mb x_o$. The corresponding algorithm is denoted as the EM-GM-GAMP. The key difference between the EM-GM-GAMP and the parametric SURE-AMP is that fitting the signal prior is an indirect adaptation for minimizing the reconstruction MSE while we directly tackle the problem by adaptively selecting the best-in-class denoiser with the least MSE. When the signal distribution can be well approximated by a GM model, fitting the prior and minimizing MSE lead to subtle difference. However, for distributions that are difficult to be approximated as the finite sum of Gaussians, as we demonstrate in Section \ref{sec:simu}, the parametric SURE-AMP algorithm provides a better solution. In terms of computational complexity, the parametric SURE-AMP significantly outperforms the EM-GM-GAMP with the linear parameterization of the denoisers.

In \cite{Kamilov2014}, the authors generalized the EM step with an adaptive prior selection function. The proposed adaptive generalized AMP (adaptive GAMP) algorithm includes the EM-GM-GAMP as a special case. Although the general form of the prior adaptation also enables other learning methods, i.e. maximum-likelihood (ML), to be deployed in the AMP framework, in principle the adaptive GAMP still focuses on fitting the signal prior rather than directly minimizing the reconstruction MSE. 

The parametric SURE-AMP framework was first introduced in the preliminary work \cite{guo2014}. In the current paper, detailed analysis of the algorithm, more kernel families and parameter optimization scheme are presented. Extensive simulation with different priors are also reported in terms of both reconstruction performance and computational efficiency. While writing this paper, we become aware of another relevant work, the denoising-based AMP (D-AMP) algorithm \cite{chris2014}. The intrinsic denoising problem within AMP iterations has also been noticed by the authors. The intuition for D-AMP is to take advantage of the rich existing literature on signal denoising to enhance the AMP algorithm. In the paper, the existing image denoising algorithm BM3D has be utilized as the denoiser in D-AMP and produced the state-of-art recovery for natural images. The authors essentially share the same understanding as us for the AMP algorithm and point out the possibility of using the SURE based estimator for denoising. 

\subsection{Structure of the paper}
The remainder of the paper is organized as follows: The parametric SURE-AMP algorithm is presented in Section \ref{sec:sure-amp}. Section \ref{sec:kernel} is devoted to introducing the construction of the SURE-based parametric denoiser class. Three types of kernel families as well as the parameter optimization scheme are discussed herein. The simulation results are summarized in Section \ref{sec:simu}. We compare both the reconstruction performance and the computational complexity of the parametric SURE-AMP algorithm with other CS algorithms. We conclude the paper in Section \ref{sec:conclude}. 


\textit{Notation:} For the rest of the paper, we use boldface capital letters e.g. $\mathbf{A}$, to represent matrices, and $\mathbf{A}^T$ to denote the transpose. We use boldface small letters like $\mathbf{x}$ to denote vectors and $x_i$ to represent its $i^{th}$ element. For a vector $\mathbf{x}\in\R^{n}$, we use $<\mathbf{x}>=\frac{1}{n}\sum_i x_i$ to represent its average.

\section{Parametric SURE-AMP Framework}
\label{sec:sure-amp}
\subsection{Parametric SURE-AMP algorithm}
We begin with a description of the parametric SURE-AMP algorithm, which extends the generic AMP iteration defined in eq. \eqref{eq:AMP}, eq. \eqref{eq:AMP2} and eq. \eqref{eq:AMP3} with an adaptive signal denoising module. The implementation of the parametric SURE-AMP algorithm is summarized in Algorithm \ref{algo:sure-amp}.
\begin{algorithm}
\caption{ : Parametric SURE-AMP}
\label{algo:sure-amp}
\begin{algorithmic}[1]
\STATE \textbf{initialization:}  $\hat{\mathbf{x}}^0 = \mathbf{0}$, $\mathbf{z}^0 = \mathbf{y}$, $c^0 = <\lVert \mb z^0\rVert^2>$
\FOR{$t =0,1,2, \cdots$}
\STATE $\bf r^t =  \hat{\mathbf{x}}^t + \boldsymbol\Phi^T\mathbf{z}^t$
\STATE $\bs\theta^t = H_t(\mb{r}^t, c^t)$
\STATE $\hat{\mathbf{x}}^{t+1} = f_t(\mb{r}^t , c^t |\bs\theta^t)$
\STATE $\nu^{t+1} =<f'_t(\mb{r}^t, c^t| \bs\theta^t)>$
\STATE $\mathbf{z}^{t+1} = \mathbf{y}-\boldsymbol\Phi \hat{\mathbf{x}}^{t+1} + \frac{1}{\gamma} \nu^{t+1}\mathbf{z}^t$
\STATE $c^{t+1} = <\lVert \mb z^{t+1}\rVert^2>$ 
\ENDFOR
\end{algorithmic}
\end{algorithm}

Most of the entities have the same interpretation as in AMP: $\mb r^t$ is the noisy version of the original signal, which can be effectively approximated as $\mb r^t \approx \mb x_o + \sqrt{c^t}\mb z^t$, $z_i\sim\N(z_i; 0 ,1)$. Here $c^t$ is the estimation of the effective noise variance. A new signal estimate $\hat{\mb x}^{t+1}$ is obtained by denoising $\mb r^t$ at each iteration. The key modification to AMP is the introduction of the parametric denoising function $f_t(\cdot|\bs\theta^t)$ and the parameter selection function $H_t(\cdot)$. Consider a class of denoising functions $\mathbb{F}(\cdot|\mb Q)$ characterized by the parameter set $\mb Q$. At each iteration, the best-in-class denoiser $f_t(\cdot|\bs\theta_t)\in \mathbb{F}(\cdot|\mb Q)$ is chosen by selecting the parameter $\bs\theta^t$ via the parameter selection function $H_t(\cdot)$. We design $H_t(\cdot)$ as a function of the noisy data $\mb r^t$ and the effective noise variance $c^t$ to close the parametric SURE-AMP iteration.


The next question is what should be the parameter selection criteria for the parametric SURE-AMP algorithm. Our fundamental reconstruction goal is to obtain a signal estimate $\hat{\mb x}$ with the minimum MSE. Theoretically speaking, we want to jointly select the denoisers across all iterations.
However, solving the joint optimization is not trivial. Based on the state evolution analysis in the subsequent section, we propose to break the joint selection into separate independent steps. Specifically, the parameter vector $\bs\theta^t$ at iteration $t$ is selected by solving 
\begin{equation}
\begin{split}
\bs\theta^t &= \argmin_{\bs\theta}\Expect[(\hat{\mb x}^{t+1}-\mb x_o)^2]\\
&=\argmin_{\bs\theta}\Expect\{[f_t(\mb r^t, c^t|\bs\theta)-\mb x_o]^2\}
\end{split}
\label{eq:argmin1}
\end{equation}
which assumes the optimality of $[\bs\theta^0, \cdots,\bs\theta^{t-1}]$ in the previous iterations. As the signal estimate $\hat{\mb x}^t$ is optimized within the denoiser class at each step, one would expect to obtain a "global" optimal reconstruction as the algorithm converges. 


\subsection{SURE based denoiser selection}
The Stein's unbiased estimate (SURE) is an unbiased estimate for MSE. It becomes more accurate as more data is available, which is particularly apt for AMP since it is designed with the large system limit in mind. It has been widely used as the surrogate for the MSE to tune the free parameters of estimation functions for signal denoising. In \cite{stein1981}, it has been proved that for the Gaussian noise corrupted signal, the calculation of SURE can be performed entirely in terms of the noisy observation. This property is summarized in the following theorem.

\begin{Theorem}\cite{stein1981}
\label{T:sure}
Let $x_o$ be the signal of interest and $r= x_o+\sqrt{c}z$ be noisy observation with $z\sim\N(z;0,1)$. Without loss of generality, we assume the denoising function $f(r,c|\bs\theta)$ is parameterized by $\bs\theta$ and has the form 
\begin{equation}
f(r, c|\bs\theta) = r + g(r,c|\bs\theta)
\end{equation} 
The denoised signal is obtained through $\hat{x}=f(r, c|\bs\theta)$. Then SURE is defined as the expected value over the noisy data alone and is the unbiased estimate of the MSE. That is, 
\begin{equation}
\begin{split}
\Expect_{\hat{x},x_o}\{ (\hat{x}-x_o)^2 \} &=\Expect_{r,x_o}\{[f(r,c|\bs\theta)-x_o]^2\}\\
& = c + \Expect_r\{g^2(r,c|\bs\theta) + 2c g'(r,c|\bs\theta)\}
\end{split}
\label{eq:MSE}
\end{equation}
\end{Theorem}
For the complete proof of Theorem \ref{T:sure} please refer to \cite{stein1981,raphan2011least}. According to Theorem \ref{T:sure}, the parameter selection for the parametric SURE-AMP algorithm can thus be conducted via the minimization of SURE. By the law of large numbers, the expectation in \eqref{eq:MSE} can be approximated as the average over multiple realizations of the noisy data $r$. For parametric SURE-AMP, we naturally have a vector $\mb r^t$ at each iteration. Since the term $c$ will disappear in the minimization of eq. \eqref{eq:MSE}, the corresponding parameter selection function is thus defined as
\begin{equation}
\begin{split}
\bs\theta^t & = H_t(\mb r^t, c^t)\\
& = \argmin_{\bs\theta}<g^2(\mb r^t, c^t|\bs\theta) + 2c^tg'(\mb r^t, c^t|\bs\theta)>
\end{split}
\label{eq:H}
\end{equation}
It fundamentally eliminates the dependency on the original signal for selecting the denoisers with the minimum MSE. Applying eq. \eqref{eq:H} into line 4 of Algorithm \ref{algo:sure-amp} we have a complete parametric SURE-AMP algorithm.

\subsection{State evolution}
\begin{figure}[t!]
\begin{minipage}[t!]{1\linewidth}
  \centering
  \centerline{\includegraphics[width=9cm]{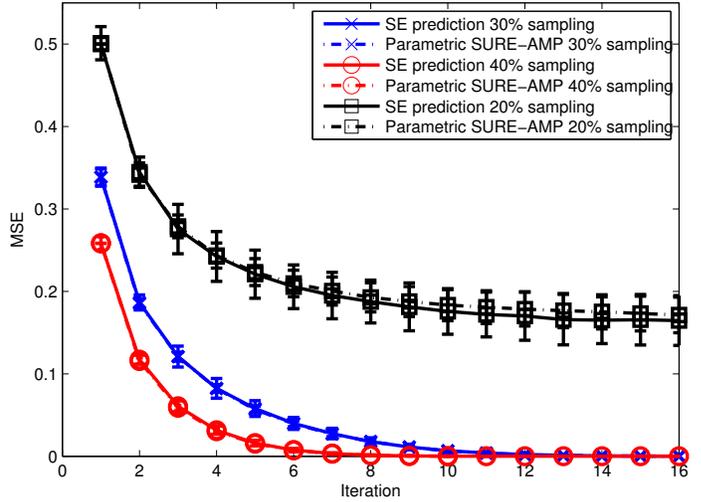}}
\end{minipage}
\vspace*{-0.3cm}
\caption{The actual MSE for the noiseless Bernoulli-Gaussian data reconstruction at each parametric SURE-AMP iteration versus the state evolution prediction. The signal is generated i.i.d according to eq. \eqref{eq:BG}. The first piecewise linear kernel family is utilized within the parametric SURE-AMP algorithm, which will be discuss in section \ref{subsec:kernel1}. The reconstruction MSE is an average over 100 Monte Carlo realizations.}
\label{fig:BGSE}
\end{figure}

One distinguishable feature of the AMP algorithm is that its asymptotic behaviour can be accurately characterized by the simple state evolution (SE) formalism in the large system limit \cite{Donoho10messagepassing,dynamicAMP}. Specifically, the SE equation can be used to predict the reconstruction MSE for AMP with a large Gaussian random measurement matrix. As an extension of the AMP algorithm, one expects the parametric SURE-AMP would also follow the SE analysis incorporating the denoising adaptation. We hereby formally summarize our finding: 
\begin{Finding}
Starting with $\tau^0=\frac{\lVert\mb y\rVert^2}{m}$, the state evolution equation for the parametric SURE-AMP algorithm has the following iterative form 
\begin{align}
\bar{\bs\theta}^t =& H_t(x+\sqrt{\tau^t}z, \tau^t)\\
\tau^{t+1} =& \sigma^2_w + \frac{1}{\gamma}\Expect\{ \tau^t f'_t(x+\sqrt{\tau^t}z, \tau^t|\bar{\bs\theta}^t)\}
\end{align}
where $x\sim p(x_o)$ has the same marginal distribution as the original signal, $z\sim\N(z;0,1)$ is the white Gaussian noise. In the large system limit, i.e. $m\rightarrow\infty$, $n\rightarrow\infty$ with $\gamma=m/n$ fixed, the MSE of parametric SURE-AMP estimate at iteration $t$ can be predicted as
\begin{equation}
\Expect\{(\mb x_o - \hat{\mb x}^t)^2\} = \sigma^2_w + \frac{1}{\gamma}\Expect\{\left[x - f_t(x+\sqrt{\tau^t}z, \tau^t|\bar{\bs\theta}^t) \right]^2 \}
\end{equation}
\label{F:1}
\end{Finding}
We use the term Finding here to emphasize the lack of rigorous proof. However, the empirical simulation supports our finding. In Fig. \ref{fig:BGSE}, the state evolution prediction for the noiseless BG signal reconstruction with the parametric SURE-AMP algorithm is compared against the Monte Carlo average at multiple iterations. It is clear from the figure that at various sampling ratios, SE accurately predicts the MSE of the parametric SURE-AMP reconstruction. 

Finding \ref{F:1} coincides with the SE analysis for the adaptive GAMP algorithm in \cite{Kamilov2014} when the output channel is assumed to be Gaussian white noise and $H_t(\cdot)$ is the prior fitting function. The authors have proved that when $H_t(\cdot)$ has the weak pseudo-Lipschitz continuous property and the denoising function $f_t(\cdot|\bs\theta^t)$ is Lipschitz continuous, the adaptive GAMP can be asymptomatically characterized by the corresponding state evolution equations in the large system limit. Unfortunately, their analysis does not apply directly to the parametric SURE-AMP algorithm since our $H_t(\cdot)$ and $f_t(\cdot|\bs\theta^t)$ do not satisfy the required pseudo-Lipschitz continuous properties. The theoretical proof of Finding \ref{F:1} is beyond the scope of this paper and remains an open question for further study.


\section{Construction of the Parametric Denoiser}
\label{sec:kernel}
The reconstruction quality of the parametric SURE-AMP algorithm primarily counts on the construction of the adaptive denoiser class and the tuning of free parameters. We choose to form the denoiser $f_t(\cdot|\bs\theta^t)$ as a weighted sum of some kernel functions to give an adaptive non-linearity. To be specific,
\begin{equation}
f_{t,i}(\mb r^t, c^t|\bs\theta^t) = \sum_{i=1}^{k} a_{t,i} f_{t,i}(\mb r^t|\bs\vartheta_{t,i}(c^t))
\end{equation}
where $f_{t,i}(\mb r^t|\bs\vartheta_{t,i}(c^t))$ is the non-linear kernel function with $\bs\vartheta_{t,i}(c^t)$ summarizes all non-linear parameters that depend on the effective noise variance $c^t$. The linear weight for the kernel function is represented with $a_i^t$. At each parametric SURE-AMP iteration, we need to optimize the parameter set $\bs\theta^t =[a_{t,i}, \bs\vartheta_{t,i}]_{i=1}^{k}$ to select the best denoising function in the class. For the rest of this section, we drop the iteration index $t$.

This parameterization method for denoisers has been used before. In \cite{raphan2011least}, the "bump" kernel family is designed to approximate the MMSE estimator of the generalized Gaussian signal. In \cite{Blu2007, Luisier2006, Unser2007}, the exponential kernels are specifically designed for natural image denoising in the transformed domain. In this section, we start by presenting three types of kernel families for both sparse and heavy-tailed signal denoising. Then we will explain the parameter optimization rule for both linear and non-linear parameters of the kernels. Finally the constructed denoiser is applied to three different signal priors to validate the design.


\subsection{Kernel families}

\begin{figure*}[t!]
\begin{minipage}[h]{.32\linewidth}
  \centering
  \centerline{\includegraphics[width=5.2cm]{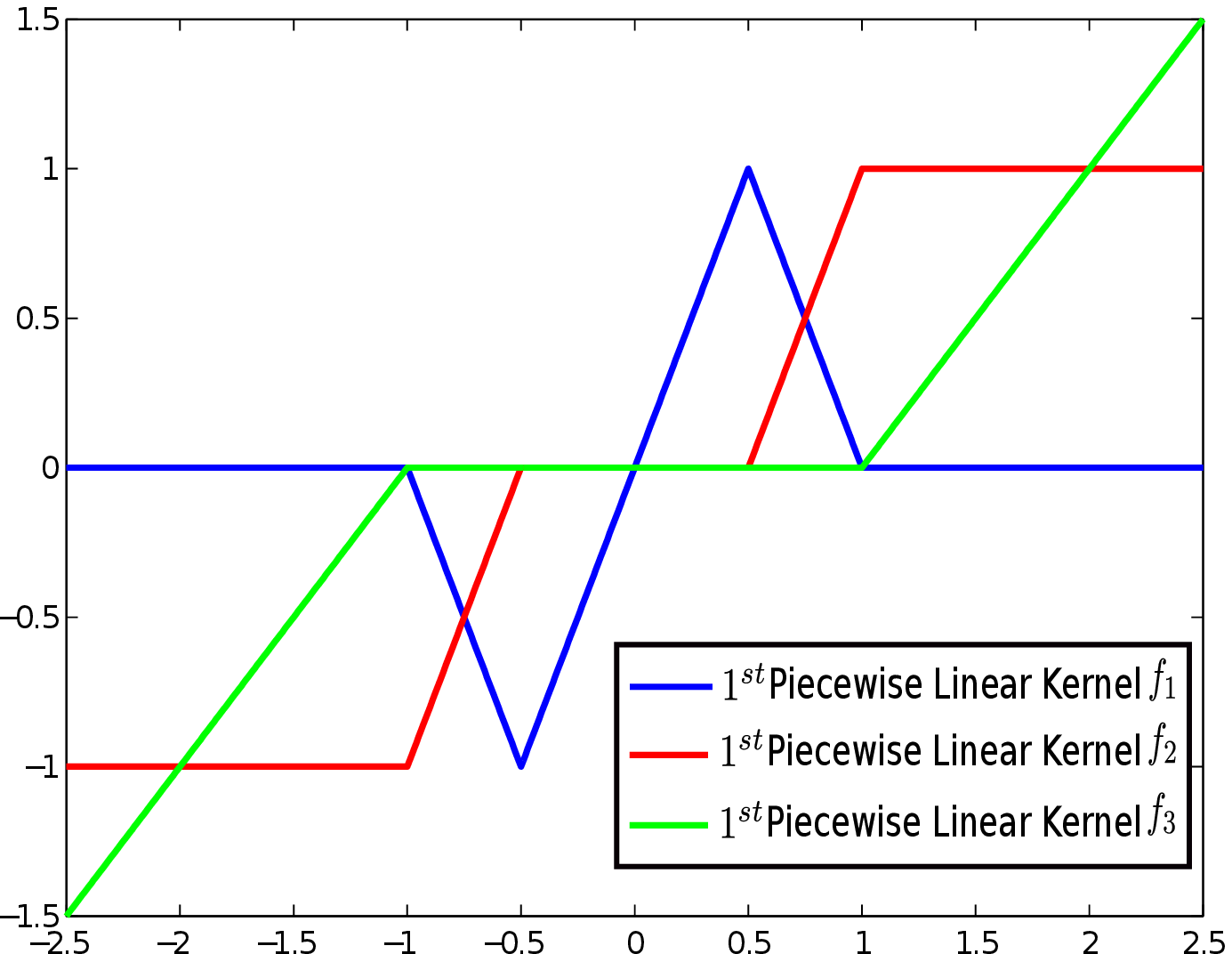}}
  \small\centerline{(a)}\medskip
\end{minipage}
\hfill
\begin{minipage}[h]{0.32\linewidth}
  \centering
  \centerline{\includegraphics[width=5.2cm]{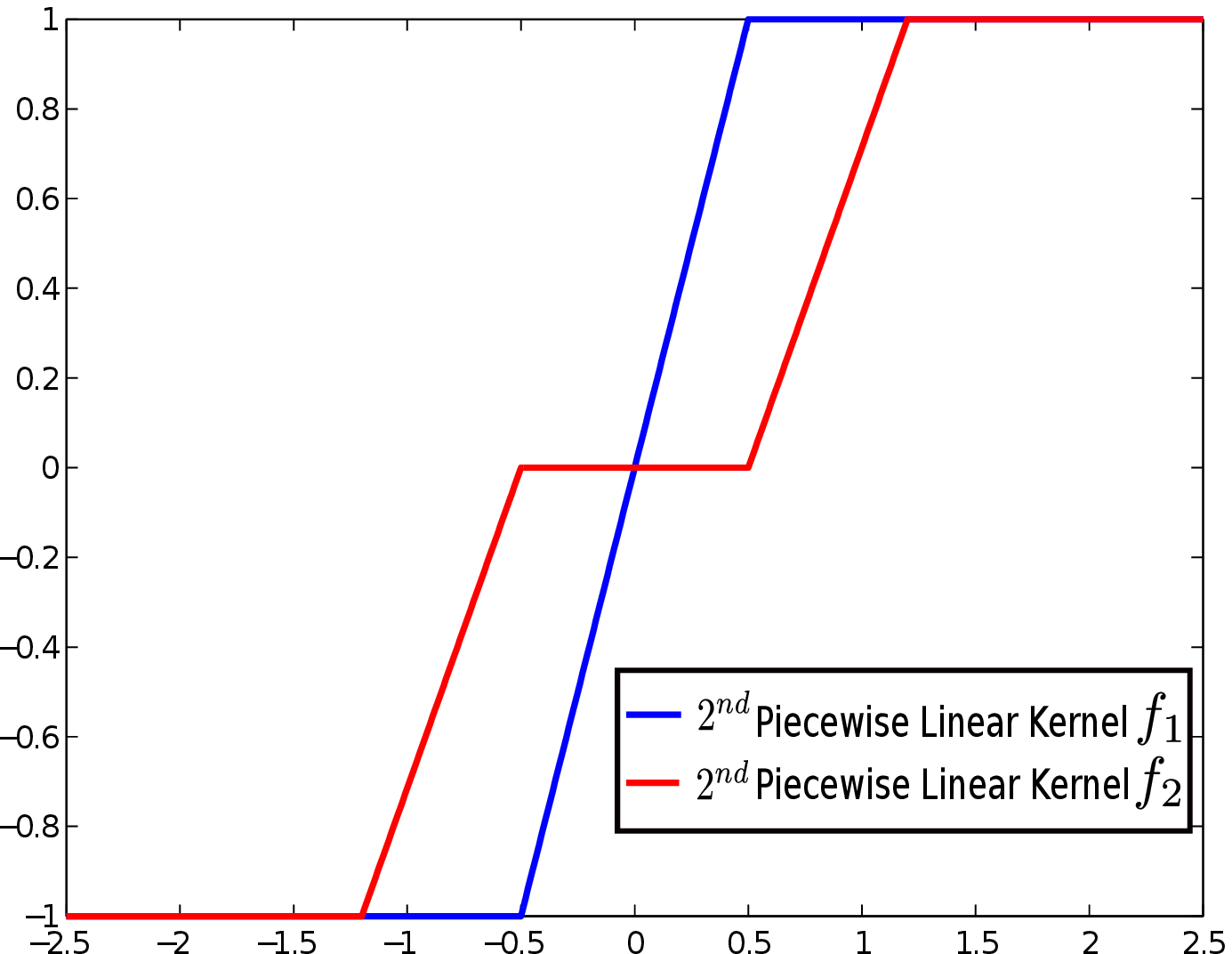}}
  \small\centerline{(b)}\medskip
\end{minipage}
\hfill
\begin{minipage}[h]{0.32\linewidth}
\centering
 \centerline{\includegraphics[width=5.2cm]{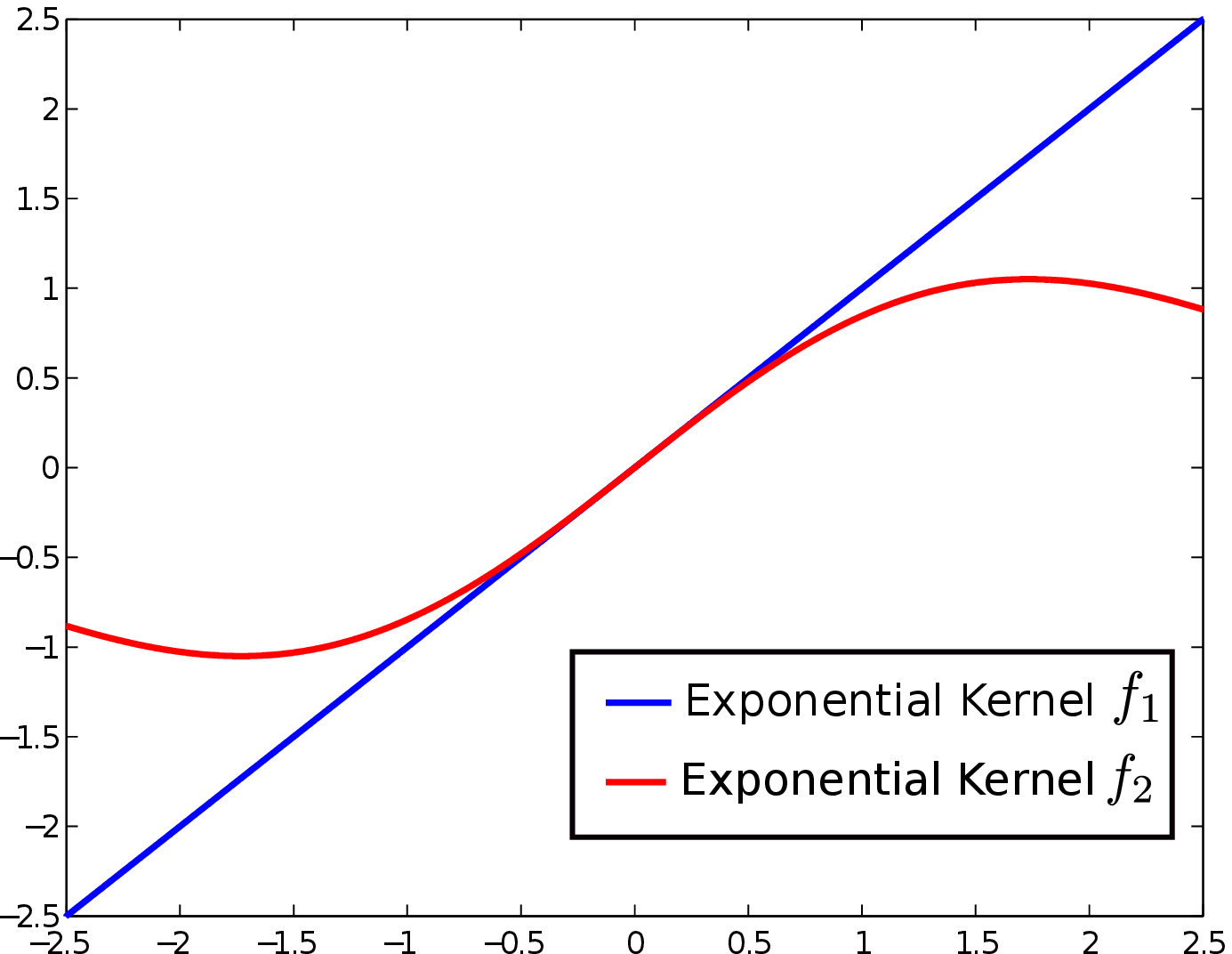}}
\small\centerline{(c)}\medskip
\end{minipage}
\vspace*{-0.3cm}
\caption{Kernel families used for linear parameterization of the SURE based denoiser: (a) the first piecewise linear kernel family (b) The second piecewise linear kernel family. (c) The exponential kernel family.}
\label{fig:kernel}
\end{figure*}
\subsubsection{First piecewise linear kernel family}
\label{subsec:kernel1}
The underlining principle for the kernel function design is to keep it simple and flexible at the same time. One way to do this is to use the piecewise linear function as the kernel format and proposed the first piecewise linear kernel family which consists three kernel functions: 
\begin{align}
\label{eq:1st1}
f_1(r|\alpha_1)&=\begin{cases}
0 & r\leq -2\alpha_1 , r \geq 2\alpha_1 \\
-\frac{r}{\alpha_1}-2 & -2\alpha_1< r < -\alpha_1 \\
\frac{r}{\alpha_1} & -\alpha_1\leq r \leq \alpha_1 \\
-\frac{r}{\alpha_1}+2 & \alpha_1< r< 2\alpha_1 \\
\end{cases}\\
\label{eq:1st2}
f_2(r|\alpha_1,\alpha_2) &= \begin{cases}
-1 & r\leq -\alpha_2 \\
\frac{r+ \alpha_1}{\alpha_2-\alpha_1} & -\alpha_2 < r < -\alpha_1 \\
0 & -\alpha_1 \leq r \leq \alpha_1 \\ 
\frac{r-\alpha_1}{\alpha_2-\alpha_1} &  \alpha_1 < r < \alpha_2 \\
1 & r\geq \alpha_2
\end{cases} \\
\label{eq:1st3}
f_3(r|\alpha_2) &= \begin{cases}
r + \alpha_2 & r\leq -\alpha_2 \\
0 & -\alpha_2 < r < \alpha_2 \\
r- \alpha_2 & r \geq \alpha_2
\end{cases}
\end{align}
where $\alpha_1>0$ and $\alpha_2>0$ are hinge points closely related to the effective noise level $c$. The three kernels are plotted in Fig. \ref{fig:kernel}(a). Eq. \eqref{eq:1st3} is the soft thresholding function to promote sparsity. It sets all vector elements whose magnitude smaller than $\alpha_2$ to zero and keeps the linear behaviour of large elements. The linear part with positive gradient in eq. \eqref{eq:1st1} aims to soften the "brutal" correction of the soft thresholding function on the small elements. It is designed for removing the Gaussian perturbation for small but non-zero elements of compressible signals. Eq. \eqref{eq:1st2} is constructed to add a denoising transition between the small and large elements to increase the denoiser flexibility. With proper rescaling of the three kernels and appropriate setting for the hinge points, we expect the denoiser class constructed with the first piecewise linear kernels to be flexible and accurate enough to capture the evolving shape of the MMSE estimators for CS signals at different noise levels.  

\subsubsection{Second piecewise linear kernel family}
In \cite{AMP}, the AMP reconstruction power for three canonical CS signal models has been demonstrated, one of which has most of the vector elements taking their the value from the discrete set $\mathbb{D}\equiv\{-\varsigma, \varsigma\}$ and the rest are real numbers from the open continuous set $\mathbb{C}\equiv(-\varsigma, \varsigma)$. We denote this signal model as the $k$-dense signal and give the explicit pdf as follows
\begin{equation}
p_{\mbox{\tiny KD}}(x) = \frac{(1-\lambda)}{2}\delta(x+\varsigma) + \frac{(1-\lambda)}{2}\delta(x-\varsigma) + \lambda\U(-\varsigma, \varsigma)
\label{eq:k-dense}
\end{equation}
where $\U$ represents the pdf of the continuous components. In the CS literature, the $k$-dense signal has been considered before as the $k$-simple signal in \cite{donoho2009counting}. The face counting theory has been established to bound the minimum sampling ratio for the perfect reconstruction of the $k$-dense signal via the convex optimization. In \cite{AMP}, the soft thresholding function with the adaptive thresholding level is suggested as a generic AMP algorithm for such signals. For the $k$-dense signals, we propose the second piecewise linear kernel functions to construct the denoiser.   
\begin{align}
\label{eq:2nd1}
f_1(r|\beta_1) &= \begin{cases}
-1 & r \leq -\beta_1 \\
\frac{r}{\beta_1} & -\beta_1<r<\beta_1 \\
1 & r \geq \beta_1
\end{cases}\\
\label{eq:2nd2}
f_2(r|\beta_1, \beta_2) &= \begin{cases}
-1 & r \leq -\beta_2 \\
\frac{r+\beta_1}{\beta_2-\beta_1} & -\beta_2<r<-\beta_1 \\
0 & -\beta_1\leq r \leq \beta_1 \\
\frac{r-\beta_1}{\beta_2-\beta_1} & \beta_1 < r < \beta_2 \\
1 & r\geq \beta_2
\end{cases}
\end{align}

Similar to the first piecewise linear kernel family, the hinge points $\beta_1$ and $\beta_2$ depend on the effective Gaussian noise level. With proper scaling of the second piecewise linear kernels, the constructed denoiser is able to mimic the MMSE estimator behaviour for the $k$-dense signal under different noise levels.  

\subsubsection{Exponential kernel family}
For the third type of kernel family, we resort to more sophisticated exponential functions.  
\begin{align}
\label{eq:ex1}
f_1(r) & = r\\
\label{eq:ex2}
f_2(r|T)& =r e^{-\frac{r^2}{2T^2}}
\end{align} 
This kernel family is motivated from the derivatives of Gaussians (DOG) and has been used for natural image denoising in the transformed domain in \cite{Blu2007, Luisier2006, Unser2007}. The virtue of DOGs is that they decay very fast and ensure a linear behaviour close to the identity for large elements \cite{Unser2007}. It has been demonstrated that with kernels defined in eq. \eqref{eq:ex1} and eq. \eqref{eq:ex2}, the constructed denoiser delivers the near-optimal performance regarding both quality and computational cost. The parameter $T$ in eq. \eqref{eq:ex2} has the same functionality as the hinge points for the piecewise linear kernels. It controls the transition between small and large elements and is linked tightly with the effective noise variance. Given the fact that most natural images are compressible in the wavelet or DCT domain, we believe that the exponential kernel family used for image denoising can also be applied in the parametric SURE-AMP algorithm to recovery compressible signals. 

One thing worth noting is that the proposed kernel families are not designed to fit any specific signal prior, but are motivated from the general sparse or compressible pattern. Thus they are, to some extent, suitable for many CS signal reconstructions. It is also straightforward to construct new kernel functions to increase the sophistication of the constructed denoiser. For the exponential kernel family, high order DOGs can be used. For the piecewise linear kernel families, more functions with various hinge points could be added. In our work, we find that with just three kernel functions, the constructed denoiser is able to deliver a near Bayesian optimal performance. Moreover, we do not necessarily require the denoiser class to contain the true MMSE estimator to achieve good reconstruction performance. As proved in \cite{Remi2011}, denoisers constructed by the piecewise linear kernels are not eligible for the true MMSE estimator since they are not $C^{\infty}(\R^n)$. Nevertheless, they exhibit excellent performance for the CS signal denoising and integrate well with the parametric SURE-AMP reconstruction as we will see later. When the parametric denoiser class includes all possible MMSE estimators for a specific prior, the parametric SURE-AMP algorithm is guaranteed to obtain the BAMP recovery.

\subsection{Non-linear parameter tuning for kernel functions}
To cope with the developing noise level during the parametric SURE-AMP iteration, the aforementioned kernel functions all have some non-linear dependency, i.e. the hinge point and the variance for the exponential kernel. While the non-linearity is necessary, finding the global optimizer for the non-linear parameter can be computationally expensive. To mitigate this problem, we propose a fixed linear relationship between the non-linear parameters and the effective noise level. Since at each parametric SURE-AMP iteration we obtain an estimated effective noise variance $c^t$, the non-linear parameters are consequently selected. In this section, we will explain the non-linear parameter tuning for all three kernel families.
\subsubsection{First piecewise linear kernel family}
In \cite{Donoho2010}, the authors discussed the thresholding choice for iterative reconstruction algorithms for compressed sensing. For iterative soft thresholding, they proposed to set the threshold as a fixed multiple of the standard derivation of the effective noise variance. The rule of thumb for the multiple is between 2 and 4. This threshold choice has been tested with the stagewise orthogonal matching pursuit (StOMP) algorithm \cite{Donoho_MAI2012} and the underling rationale has been explained therein. For the first piecewise linear kernel family which has the soft thresholding aspect, we take their recommendation and set the hinge points as
\begin{equation}
\alpha_1 = 2\sqrt{c}, \quad \alpha_2 = 4\sqrt{c}
\label{eq:alpha}
\end{equation}

\subsubsection{Second piecewise linear kernel family}
In \cite{Norbert2014}, a novel iterative dense recovery (IDR) algorithm is proposed to replace the MMSE estimator for the $k$-dense signal with an adaptive denoiser within the AMP iteration. The essence of the IDR is the employment of a piecewise linear function with one flexible hinge point to approximate the MMSE estimator class. Inspired by the selection of hinge point in \cite{Norbert2014}, we choose to fix the linear ratio for the second piecewise linear kernels as following
\begin{equation}
\label{eq:2ndr}
\beta_1=\frac{1}{1+6\sqrt{c}},\quad \beta_2=\frac{1}{1+2\sqrt{c}}
\end{equation}
The ratio in \eqref{eq:2ndr} is based on the empirical denoising experiments with $k$-dense signals under different noise level. Although not very critical, we find it to be a good choice for implementing the parametric SURE-AMP algorithm to recover the $k$-dense signal.
\subsubsection{Exponential kernel family}
For the non-linear parameter of the exponential kernel, we adopt the recommendation in \cite{Unser2007} and set $T$ as 
\begin{equation}
\label{eq:T}
T = 6\sqrt{c}
\end{equation}

It has been demonstrated through extensive simulations in \cite{Unser2007} that the image denoising quality is not very sensitive to the ratio between $T$ and $\sqrt{c}$. Eq. \eqref{eq:T} is shown to be a practical setting for removing various noise perturbation irrespective of the images. The denoising and reconstruction simulations in the subsequent sections will also confirm that it is a plausible choice for both sparse and heavy-tailed signals. 


\subsection{Linear parameter optimization}
With the non-linear parameters fixed with the effective noise variance, the only parameters left to be optimized are the kernel weights $a_i$. Denote $\varepsilon$ as the MSE of the denoised signal using the parametric function $f(\mb r,c|\bs\theta)$. With Theorem \ref{T:sure}, we have 
\begin{equation}
\varepsilon = c + <g^2(\mb r, c|\bs\theta) + 2cg'(\mb r,c|\bs\theta)>
\end{equation}  
where 
\begin{equation}
\begin{split}
g(\mb r, c|\bs\theta)=& f(\mb r,c|\bs\theta)-\mb r\\
=&\sum_{i=1}^{k}a_i f_i(\mb r|\bs\vartheta_i(c))-\mb r
\end{split}
\end{equation} 
Optimizing the weights $a_i$ to achieve the minimum MSE requires differentiation of $\varepsilon$ over $a_i$ and solving for all $i\in(1,\cdots,k)$. 
\begin{equation}
\begin{split}
& \frac{d\varepsilon}{d a_i}=<2g(\mb r,c|\bs\theta)\frac{d}{d a_i}g(\mb r,c|\bs\theta) + c\frac{d}{d a_i}g'(\mb r|\bs\theta)>=0\\
\Longleftrightarrow & \sum_{j=1}^{k}<a_j f_j(\mb r|\bs\vartheta_j(c))f_i(\mb r|\bs\vartheta_i(c))>=-c <f'_i(\mb r|\bs\vartheta_i(c))>
\end{split}
\end{equation}
All equations can be summarized in the following matrix form
\begin{equation}
\label{eq: ParameterOp}
\underbrace{\begin{bmatrix}
<f^2_1> & \cdots  & <f_1f_k>\\
\vdots & \ddots & \vdots \\
<f_kf_1> & \cdots & <f_k^2>
\end{bmatrix}}_{\mathcal{F}}\underbrace{\begin{bmatrix}
a_1\\
\vdots\\
a_k
\end{bmatrix}}_{\mathcal{A}}=-c\underbrace{\begin{bmatrix}
<f'_1>\\
\vdots\\
<f'_k>
\end{bmatrix}}_{\mathcal{D}}
\end{equation}
The linear system can then be solve by 
\begin{equation}
\label{eq:AFC}
\mathcal{A}=-c\mathcal{F}^{-1}\mathcal{D}
\end{equation}

In summary, the linear kernel weights can be easily optimized by solving a linear system of equations. We will demonstrate later that this linear parameterization is very advantageous in terms of the computational complexity.  
\subsection{Denoising performance}
To validate our proposed kernel families and the parameter optimization scheme, we compare the optimized parametric denoisers alongside with the MMSE estimator for BG and $k$-dense signals.
\begin{figure}[t!]
\begin{minipage}[t!]{1\linewidth}
  \centering
  \centerline{\includegraphics[width=9cm]{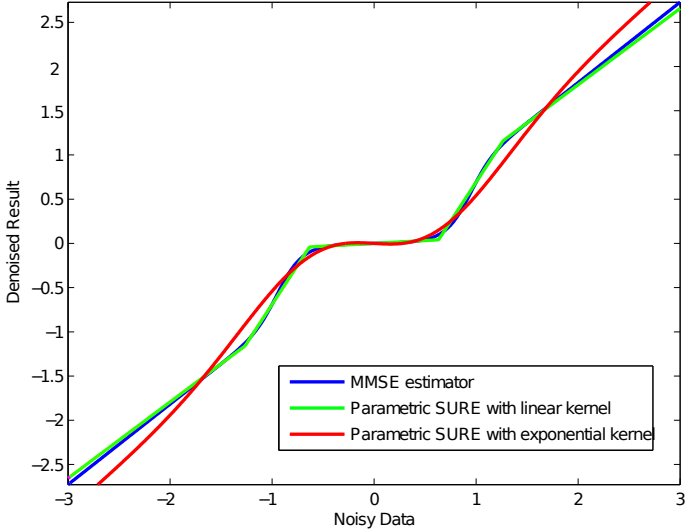}}
\end{minipage}
\vspace*{-0.3cm}
\caption{MMSE estimator and parametric SURE for the noisy Bernoulli-Gaussian data. The noise variance $c$ is 0.1. The reconstruction error for the MMSE estimator, the SURE estimator with the first piecewise linear kernel and the SURE estimator with the exponential kernel are $0.020615$, $0.020788$ and $0.022047$, respectively.}
\label{fig:BG_denoise}
\end{figure}

In Fig. \ref{fig:BG_denoise} we can see that with just three kernel functions from the first piecewise linear kernel family and the suggested parameter optimization in eq., the constructed denoiser achieves an excellent agreement with the MMSE estimator for the noisy BG data. The MSE difference between the denoised signal using the SURE based parametric denoier and the Bayesian optimal denoising is neglectable. The exponential kernel family also does a good job at capturing the key structure of the MMSE estimator, especially in the vicinity of small values where most of the data concentrates. 

\begin{figure}[t!]
\begin{minipage}[t!]{1\linewidth}
  \centering
  \centerline{\includegraphics[width=9cm]{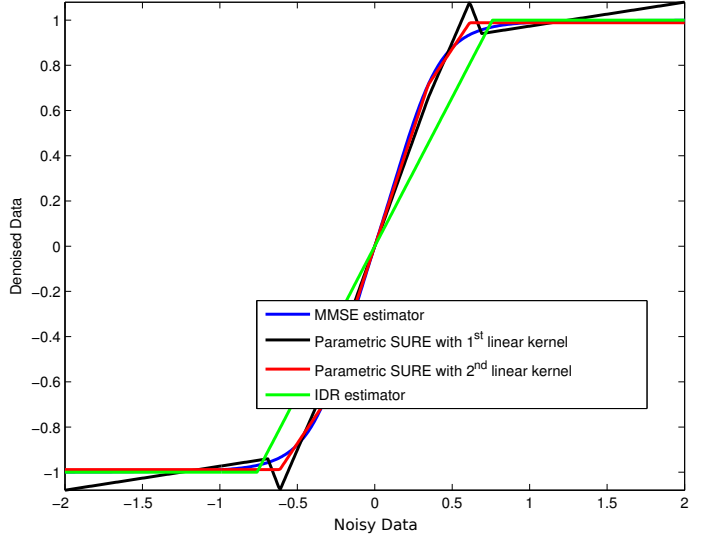}}
\end{minipage}
\vspace*{-0.3cm}
\caption{MMSE estimator and parametric SURE for the noisy $k$-dense data. The noise variance $c$ is 0.1. The reconstruction error for the MMSE estimator, the SURE with the second piecewise linear kernel, the SURE estimator with the first piecewise linear kernel and the IDR denoiser are $0.0243$ and $0.0248$, $0.0251$ and $0.0315$ respectively.}
\label{fig:debse_denoise}
\end{figure}

In Fig. \ref{fig:debse_denoise} we compare the MMSE estimator, the SURE based parametric denoisers with the proposed two piecewise linear kernel families, and the IDR estimator \cite{Norbert2014} for the $k$-dense signal denoising. As demonstrated in the plot, the denoiser constructed with the second piecewise linear kernel fits the MMSE estimator better because the kernels are tailored to the $k$-dense structure. The first piecewise linear kernel based denoiser performs slightly worse  because of the unbounded $f_3$ in eq. \eqref{eq:1st3}. The IDR denoiser is a piecewise linear function with just one hinge point. Thus it misses the subtle transition between the small and large elements and performs the worst among the three. 

\begin{table*}
\centering
\begin{tabular}{|| c || c | c | c | c | c | c| c ||}
\hline
Effective noise level $c$ & 0.01 & 0.1 & 1 & 5 & 10 & 50 & 100 \\
\hline
\hline
MMSE estimator for 4-state GM & 9.9655e-3 & 0.0958 & 0.7285 & 2.1788 & 3.2088 & $\mathbf{6.5801}$ & 8.6543 \\
\hline
Exponential kernel denoiser & 9.9948e-3 & 0.0967 & 0.7200 & $\mathbf{2.1504}$ & $\mathbf{3.1606}$& 6.9979 & 9.6347 \\
\hline
Piecewise linear kernel denoiser & $\mathbf{9.9383e-3}$ & $\mathbf{0.0955}$ & $\mathbf{0.7191}$ & 2.1560 & 3.1764 & 6.6554 & $\mathbf{8.6245}$\\ 
\hline
\end{tabular}
\vspace{0.1cm}
\caption{Denoising comparison for noisy Student's-t signal with various denoisers}
\label{T:student_denoise}
\end{table*}

To check the denoising power of the proposed kernel families for heavy tailed signals, we present the averaged MSE for the Student's-t signal denoising in Table \ref{T:student_denoise}. Since there is not an explicit form for the MMSE estimator for the Student's-t prior, we compare the SURE based parametric denoiser with the GM model based denoiser, which is the MMSE estimator for the 4-state GM distribution used to approximate the Student's-t distribution. It essentially is the key denoisng approach implemented by the EM-GM-GAMP algorithm. Each figure reported in Table \ref{T:student_denoise} is an average over 100 iterations.  The SURE based denoiser with the exponential kernel and the first piecewise linear kernel both deliver similar denoising performance as the MMSE estimator for the 4-state GM approximation, if not better. This implies that the corresponding parametric SURE-AMP algorithm should be competitive with the EM-GM-GAMP for the Student's-t signal reconstruction. 



\section{Numerical Results}
\label{sec:simu}
In this section, the reconstruction performance and computational complexity of the parametric SURE-AMP algorithm, using the three types of kernel families introduced in Section \ref{sec:kernel}, are compared with other CS reconstruction algorithms. In particular, we experiment with the Bernoulli-Gaussian, $k$-dense and Student's-t signals to demonstrate the reconstruction power and efficiency of the parametric SURE-AMP algorithm. 
\subsection{Noisy signal recovery}
We first present the reconstruction quality for noisy signal recovery. For all simulations, we fixed the signal dimension to $n=10000$. Each numerical point in the plots is an average of 100 Monte Carlo iterations. To have a fair comparison, the noise level is defined in the measurement domain and quantified as 
\begin{equation}
SNR_y = 10\log_{10}\frac{\lVert\bs\Phi\mb x_o\rVert^2_2}{\lVert \mb w\rVert_2^2}
\end{equation}
The reconstruction quality is evaluated in terms of the signal to noise ratio in the signal domain, defined as
\begin{equation}
SNR_x = 10\log_{10}\frac{\lVert\mb x_o\rVert^2_2}{\lVert \mb x_o-\hat{\mb x}\rVert_2^2}
\end{equation}
The elements of the measurement matrix $\bs\Phi$ are drawn i.i.d from $\N(\Phi_{ij};0,m^{-1})$ and the matrix columns are normalized to one. For all reconstruction algorithms, the convergence tolerance is set as $10^{-6}$. The maximum iteration number is set as 100.
\subsubsection{Bernoulli-Gaussian prior}
\begin{figure}[t!]
\begin{minipage}[t!]{1\linewidth}
  \centering
  \centerline{\includegraphics[width=9.2cm]{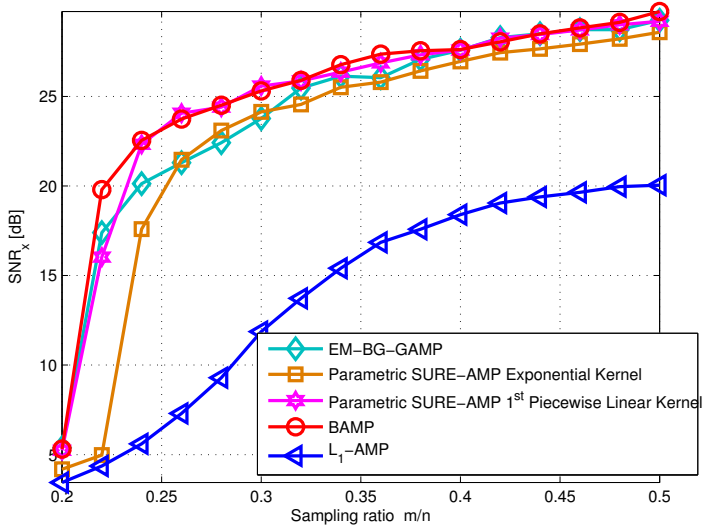}}
\end{minipage}
\vspace*{-0.3cm}
\caption{SNRx versus sampling ratio for CS recovery of noisy Bernoulli-Gaussian data.}
\label{fig:BG}
\end{figure}

The Bernoulli-Gaussian signals for the simulation are draw i.i.d from 
\begin{equation}
p(x_o) = 0.1\N(x_o;0,1) + 0.9\delta(x_0)
\label{eq:BG}
\end{equation}
We choose the noise level to be $SNR_y= 25$ dB. For comparison, we show the performance of the parametric SURE-AMP algorithm with both first piecewise linear kernel and the exponential kernel family, the EM-BG-GAMP algorithm \footnote{A special case of the EM-GM-GAMP algorithm which approximates the signal prior with a mixture of Bernoulli and Gaussian distributions. We use the implementation from http://www2.ece.ohio-state.edu/~vilaj/EMGMAMP/EMGMAMP.html.}, the $\lone$-AMP algorithm \footnote{We use the implementation from http://people.epfl.ch/ulugbek.kamilov.} and the genie BAMP \footnote{The true signal prior $p(\mb x_o)$ is assumed known for the BAMP reconstruction. It is served as the upper bound for $SNR_x$.} algorithm. The reconstruction quality $SNR_x$ for various sampling ratios are illustrated in Fig. \ref{fig:BG}.
It is obvious that the parametric SURE-AMP algorithm with the first piecewise linear kernel exhibits the near-optimal construction: for $\gamma\geq 0.24$, the difference between the parametric SURE-AMP algorithm which is blind to the signal prior and the genie BAMP algorithm is negligible. It also adequately demonstrates that SURE is a perfect surrogate for the MSE measure and the intrinsic signal property can be effectively exploited by the SURE-based denoiser. Moreover, it shows again that the proposed hinge point selection strategy in \eqref{eq:alpha} works very well regardless of the effective noise level. Compared with the EM-BG-GAMP algorithm, it delivers roughly 2 dB better recovery for $0.24\leq\gamma \leq0.3$. For $\gamma>0.36$, EM-BG-GAMP also delivers reconstruction performance that is very close to the genie BAMP result. It is because the kernels used in EM-BG-GAMP to fit the data are essentially the prior for generating the data. For the parametric SURE-AMP algorithm with the exponential kernels, it is roughly 1 dB worse than its counterpart with the first piecewise linear kernel and the Bayesian optimal reconstruction for $\gamma\geq 0.26$. This comes as no surprise as we have already seen in Fig. \ref{fig:BG_denoise} that the denoiser based on exponential kernels doesn't capture the MMSE estimator structure for data with large magnitude. Nevertheless, it still demonstrates significant improvement over the $\lone$-AMP reconstruction for which no statistical property of the original signal is exploited. 

\subsubsection{k-Dense signal}
In \cite{Norbert2014}, extensive simulation has been conducted to compare the IDR algorithm performance with the state-of-art algorithms for the noisy $k$-dense signal reconstruction. Thus in this paper, we use the same setting and mainly compare the parametric SURE-AMP using two piecewise linear kernel families with the IDR, EM-GM-GAMP and the genie BAMP algorithm. The $k$-dense signal is generated i.i.d from eq. \eqref{eq:k-dense} with $\lambda =0.1$ and $\U$ being the uniform distribution. The noise level is $SNR_y = 28$ as in \cite{Norbert2014}. For the EM-GM-GAMP algorithm we found that as the number of Gaussian components increase, the reconstruction quality gets better. Thus we used 20 Gaussian mixture to fit the $k$-dense prior, which is the largest number allowed for the EM-GM-GAMP MATLAB package. 
\begin{figure}[t!]
\begin{minipage}[t!]{1\linewidth}
  \centering
  \centerline{\includegraphics[width=8.7cm]{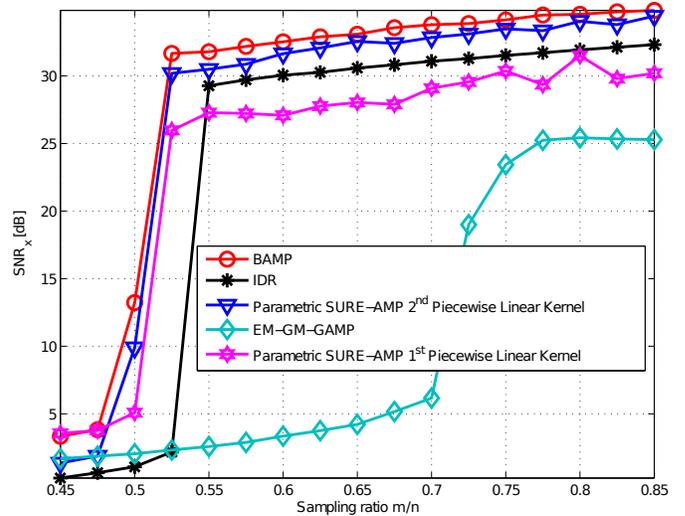}}
\end{minipage}
\vspace*{-0.3cm}
\caption{SNRx versus sampling ratio for CS recovery of noisy $k$-dense data.}
\label{fig:dense}
\end{figure}
The parametric SURE-AMP with the second piecewise linear kernel is only slightly worse than the genie BAMP reconstruction. There is roughly $0.5$ dB difference between the two for $\gamma>0.5$. Comparing to the IDR reconstruction, there is a consistent 2 dB improvement for $\gamma\geq 0.55$. This reconstruction quality gain is predictable as we have already demonstrated in the denoising section in Fig. \ref{fig:debse_denoise}. It is also reasonable that the first piecewise linear kernel based parametric SURE-AMP does not perform as well as IDR and the second piecewise linear kernel. It is in general 2 dB worse than the IDR and 5 dB worse than the genie BAMP bench mark. This is mainly because the first piecewise linear kernel fails to correct the large coefficients to be $\pm 1$. However, it still greatly outperforms the EM-GM-GAMP algorithm. The failure of the EM-GM-GAMP in this case is probably because the algorithm gets stuck at some local minima when fitting the prior. This example confirms the advantageous motivation for the parametric SURE-AMP algorithm: minimizing the MSE is the direct approach to obtain the best reconstruction.   
\subsubsection{Student-t prior}
\begin{figure}[t!]
\begin{minipage}[t!]{1\linewidth}
  \centering
  \centerline{\includegraphics[width=9.4cm]{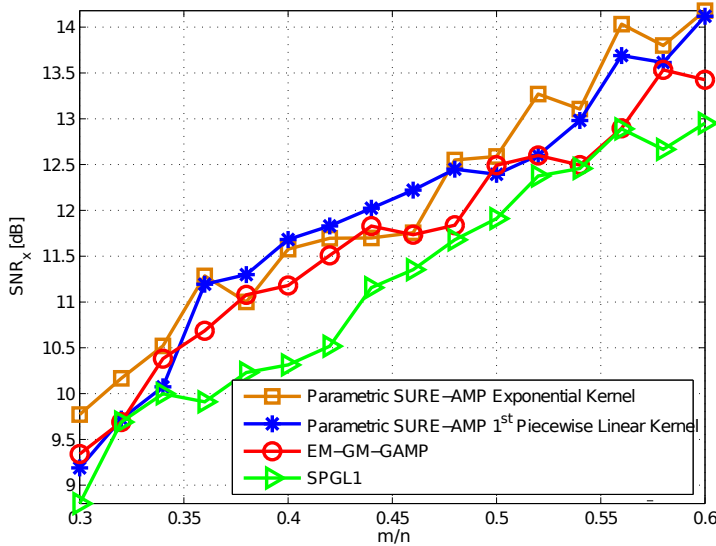}}
\end{minipage}
\vspace*{-0.3cm}
\caption{SNRx versus sampling ratio for CS recovery of noisy student-t data.}
\label{fig:student}
\end{figure}

To investigate the parametric SURE-AMP performance for signals that are not strictly sparse, we consider the Student's-t prior as a heavily-tailed distribution example. The signal is draw i.i.d according to the following distribution.
\begin{equation}
p_{\mbox{\tiny T}}(x_o) = \frac{\Gamma((q+1)/2)}{\sqrt{q\pi}\Gamma(q/2)}(1+x_o^2)^{-(q+1)/2}
\end{equation}
where $q$ controls the distribution shape. It has been demonstrated in \cite{gribonval_compressible11} that the Student's-t distribution is an excellent model to capture the statistical behaviour of the DCT coefficients for natural images. In the simulation, we set $q=1.67$, $SNR_y = 25$ dB as in \cite{Vila2013}. The parametric SURE-AMP using both exponential and the first piecewise linear kernel family are compared with the EM-GM-GAMP algorithm and LASSO via SPGL1\footnote{We run the SPGL1 in the "BPDN" mode. The MATLAB package can be found in http://www.cs.ubc.ca/labs/scl/spgl1.} \cite{BergFriedlander2008}. As we can see from Fig. \ref{fig:student}, the parametric SURE-AMP and EM-GM-GAMP have the similar reconstruction performance. This can be expected from the denoising comparison in the previous section. None of them achieves significant improvement over the $\lone$-minimization approach though. It probably because the signal is not very compressible. With more sophisticated kernel design we might achieve better performance with the parametric SURE-AMP algorithm.

\subsection{Runtime comparison}
\begin{figure}[t!]
\begin{minipage}[t!]{1\linewidth}
  \centering
  \centerline{\includegraphics[width=9cm]{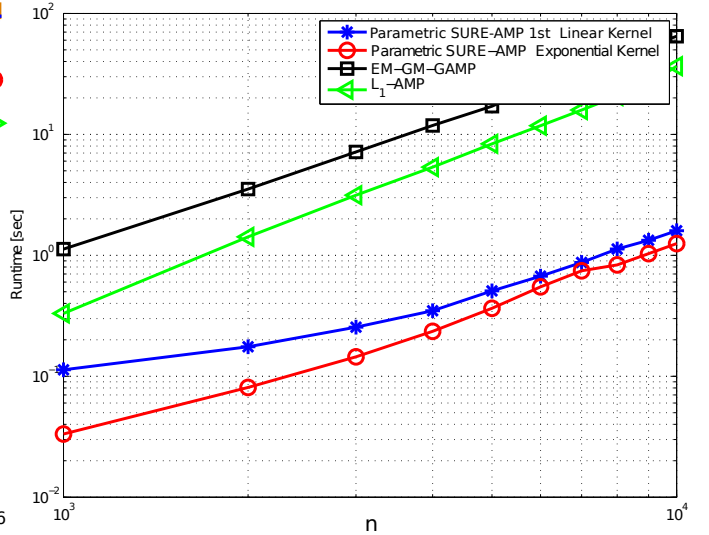}}
\end{minipage}
\vspace*{-0.3cm}
\caption{Runtime versus signal dimension for CS recovery of noisy Bernoulli-Gaussian data.}
\label{fig:BGt}
\end{figure}

\begin{figure}[t!]
\begin{minipage}[t!]{1\linewidth}
  \centering
  \centerline{\includegraphics[width=9cm]{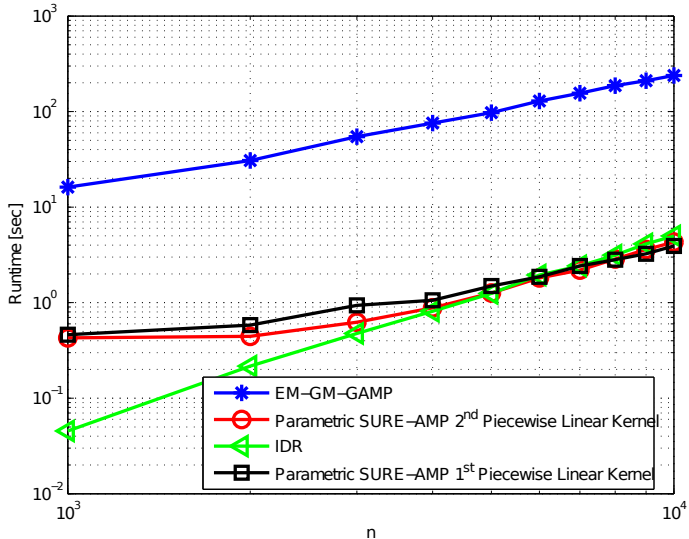}}
\end{minipage}
\vspace*{-0.3cm}
\caption{Runtime versus signal dimension for CS recovery of noisy $k$-dense data.}
\label{fig:denseT}
\end{figure}

\begin{figure}[t!]
\begin{minipage}[t!]{1\linewidth}
  \centering
  \centerline{\includegraphics[width=9cm]{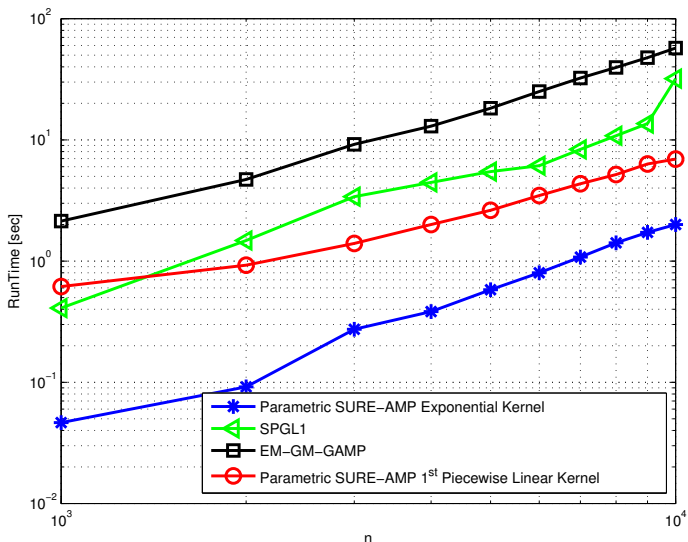}}
\end{minipage}
\vspace*{-0.3cm}
\caption{Runtime versus signal dimension for CS recovery of noisy student-t data.}
\label{fig:studentT}
\end{figure}

The parametric SURE-AMP algorithm with the simple kernel functions and linear parameterization does not only achieve the near optimal reconstruction. More importantly, it significantly reduces the computational complexity. The authors in \cite{Vila2013} has compared the EM-GM-GAMP algorithm with most of the existing CS algorithms that are blind of the prior and proved EM-GM-GAMP is the most efficient among them all. Thus in this section, we will use the EM-GM-GAMP runtime performance as the bench mark to evaluate the efficiency of the parametric SURE-AMP algorithm. For this, we fixed $\gamma = 0.5$, $SNR_y = 25$ dB and varied the signal length $n$ from 10000 to 100000. For the EM-GM-GAMP algorithm, we set the EM tolerance to $10^{-5}$ and the maximum EM iterations to $20$. The runtime for noisy recovery of the BG, $k$-dense and Student's-t data are plotted in Fig. \ref{fig:BGt}, Fig. \ref{fig:denseT} and Fig. \ref{fig:studentT} respectively. Every point in the plots is an average over 100 realizations. The algorithms tested here are the same as described before.

As with the EM-GM-GAMP algorithm, the major computational cost for the parametric SURE-AMP comes from the matrix multiplication of the vector with the measurement matrix $\bs\Phi$ and $\bs\Phi^T$ at each iteration. However, we observed a dramatic runtime improvement across all tested signal lengths for three signal priors. The parametric SURE-AMP is more than 20 times faster than the EM-GM-GAMP scheme. The algorithm efficiency can be attributed to the simple form of the kernel functions, the linear parameterization of the SURE-based denoiser and the reduced number of iterations. Consider the runtime comparison of the BG data reconstruction. The total number of the EM-BG-GAMP iterations is roughly twice as many as that of the parametric SURE-AMP algorithm. Moreover, the per-iteration computational cost is much more expensive for EM-BG-GAMP since fitting the signal prior requires many EM iterations. While for each parametric SURE-AMP iteration, only one linear system needs to be solved to optimize the adaptive estimator. When compared with the $\lone$-AMP, the runtime for each parametric SURE-AMP iteration is approximately the same. The improved runtime performance here comes from the effective denoising so that fewer iterations are required for the parametric SURE-AMP to converge. The best runtime performance of the IDR algorithm for the $k$-dense data in Fig. \ref{fig:denseT} is understandable since it only applies an adaptive thresholding function at each iteration and has no parameter optimization procedure. 


\section{Conclusion}
\label{sec:conclude}
In this paper, the parametric SURE-AMP is presented as a novel compressed sensing algorithm, which directly minimizes the MSE of the recovered signal at each iteration. Motivated from the fact that the AMP can be cast as an iterative Gaussian denoising algorithm, we propose to utilize the adaptive SURE based parametric denoiser within the AMP iteration. The optimization of the parameters is achieve by minimizing the SURE, which is an unbiased estimate of the MSE. More importantly, the minimization of SURE depends purely on the noisy observation, which in the large system limit fundamentally eliminates the need of the signal prior. This is also the first time that it has been employed for the CS reconstruction. The parametric SURE-AMP with the proposed three kernel families have demonstrated almost the same reconstruction quality as the BAMP algorithm, where the true signal prior is provided. It also outperforms the EM-GM-GAMP algorithm in terms of the computational cost. Direction for further research would involve considering other type of kernel families and the rigorous proof for the state evolution dynamics.  
\ifCLASSOPTIONcaptionsoff
  \newpage
\fi

\bibliographystyle{IEEEtran}
{\footnotesize \bibliography{SAtreeBib.bib}}

\begin{thebibliography}{10}
\providecommand{\url}[1]{#1}
\csname url@samestyle\endcsname
\providecommand{\newblock}{\relax}
\providecommand{\bibinfo}[2]{#2}
\providecommand{\BIBentrySTDinterwordspacing}{\spaceskip=0pt\relax}
\providecommand{\BIBentryALTinterwordstretchfactor}{4}
\providecommand{\BIBentryALTinterwordspacing}{\spaceskip=\fontdimen2\font plus
\BIBentryALTinterwordstretchfactor\fontdimen3\font minus
  \fontdimen4\font\relax}
\providecommand{\BIBforeignlanguage}[2]{{%
\expandafter\ifx\csname l@#1\endcsname\relax
\typeout{** WARNING: IEEEtran.bst: No hyphenation pattern has been}%
\typeout{** loaded for the language `#1'. Using the pattern for}%
\typeout{** the default language instead.}%
\else
\language=\csname l@#1\endcsname
\fi
#2}}
\providecommand{\BIBdecl}{\relax}
\BIBdecl

\bibitem{guo2014}
C.~Guo and M.~Davies, ``Bayesian optimal compressed sensing wwithout priors:
  parametric sure aapproximate message passing,'' Sept. 2014, accepted by
  European Signal Process. Conf. (EUSIPCO), Lisbon, Portugal.

\bibitem{AMP}
D.~Donoho, A.~Maleki, and A.~Montanari, ``Message-passing algorithms for
  compressed sensing,'' \emph{Proc. of the Nat. Academy of Sciences}, vol. 106,
  no.~45, pp. 18\,914--18\,919, 2009.

\bibitem{DonohoanalysisAMP}
A.~Maleki and A.~Montanari, ``Analysis of approximate message passing
  algorithm,'' in \emph{44th Ann. Conf. on Inform. Sci. and Syst. (CISS)},
  March 2010, pp. 1--7.

\bibitem{Donoho10messagepassing}
D.~Donoho, A.~Maleki, and A.~Montanari, ``Message passing algorithms for
  compressed sensing: {II}. analysis and validation,'' in \emph{IEEE Inform.
  Theory Workshop}, 2010.

\bibitem{BayAMP}
------, ``Message passing algorithms for compressed sensing: I. motivation and
  construction,'' in \emph{IEEE Inf. Theory Workshop (ITW)}.\hskip 1em plus
  0.5em minus 0.4em\relax Dublin, Ireland, 2010, pp. 1--5.

\bibitem{Donoho1995}
D.~Donoho and I.~Johnstone, ``Adapting to unknown smoothness via wavelet
  shrinkage,'' \emph{J. American Stat. Assoc.}, vol.~90, pp. 1200--1224, 1995.

\bibitem{Pesquet1997}
J.~Pesquet and D.~Leporini, ``A new wavelet estimator for image denoising,'' in
  \emph{6th IEEE Int. Conf. on Image Process. and its Applicat.}, 1997, pp.
  249--253.

\bibitem{Benazza2005}
A.~Benazza-Benyahia and J.~C. Pesquet, ``Building robust wavelet estimators for
  multicomponent images using stein's principle,'' \emph{IEEE Trans. Image
  Proc.}, vol.~14, pp. 1814--1830, 2005.

\bibitem{Luisier2006}
F.~Luisier, T.~Blu, and M.~Unser, ``Sure-based wavelet thresholding integrating
  inter-scale dependencies,'' in \emph{2006 IEEE Int. Conf. on Image Process.},
  Oct 2006, pp. 1457--1460.

\bibitem{Unser2007}
------, ``A new sure approach to image denoising: Interscale orthonormal
  wavelet thresholding,'' \emph{IEEE Trans. on Image Process.}, vol.~16, no.~3,
  pp. 593--606, March 2007.

\bibitem{Blu2007}
T.~Blu and F.~Luisier, ``The sure-let approach to image denoising,'' \emph{IEEE
  Trans. on Image Process.}, vol.~16, no.~11, pp. 2778--2786, Nov 2007.

\bibitem{raphan2011least}
M.~Raphan and P.~Simoncelli, ``Least squares estimation without priors or
  supervision,'' \emph{Neural computation}, vol.~23, no.~2, pp. 374--420, 2011.

\bibitem{stein1981}
C.~Stein, ``Estimation of the mean of a multivariate normal distribution,''
  \emph{The annals of Statistics}, pp. 1135--1151, 1981.

\bibitem{Vila2011}
J.~Vila and P.~Schniter, ``Expectation-maximization bernoulli-gaussian
  approximate message passing,'' in \emph{2011 Conf. Record of the Forty Fifth
  Asilomar Conf. on Signals, Syst. and Comput. (ASILOMAR)}, Nov 2011, pp.
  799--803.

\bibitem{Vila2013}
------, ``Expectation-maximization gaussian-mixture approximate message
  passing,'' \emph{IEEE Trans. on Signal Process.}, vol.~61, no.~19, pp.
  4658--4672, Oct 2013.

\bibitem{MagicMatrix}
\BIBentryALTinterwordspacing
F.~Krzakala, M.~M\'ezard, F.~Sausset, Y.~Sun, and L.~Zdeborov\'a,
  ``Statistical-physics-based reconstruction in compressed sensing,''
  \emph{Phys. Rev. X}, vol.~2, pp. 021\,005(1--18), May 2012. [Online].
  Available: \url{http://link.aps.org/doi/10.1103/PhysRevX.2.021005}
\BIBentrySTDinterwordspacing

\bibitem{MagicMatrixExtend}
------, ``Probabilistic reconstruction in compressed sensing: Algorithms, phase
  diagrams, and threshold achieving matrices,'' \emph{J. Stat. Mech.}, vol.
  P08009, Aug. 2012.

\bibitem{Kamilov2014}
U.~Kamilov, S.~Rangan, A.~K. Fletcher, and M.~Unser, ``Approximate message
  passing with consistent parameter estimation and applications to sparse
  learning,'' \emph{IEEE Trans on Inform. Theory}, pp. 2969--2985, 2014.

\bibitem{chris2014}
C.~A. Metzler, A.~Maleki, and R.~G. Baraniuk, ``From denoisign to compressed
  sensing,'' arXiv:1406.4175v3 [cs.IT], July 2014.

\bibitem{dynamicAMP}
M.~Bayati and A.~Montanari, ``The dynamics of message passing on dense graphs,
  with applications to compressed sensing,'' \emph{IEEE Trans. on Inf. Theory},
  vol.~57, no.~2, pp. 764--785, Feb. 2011.

\bibitem{donoho2009counting}
D.~Donoho and J.~Tanner, ``Counting faces of randomly projected polytopes when
  the projection radically lowers dimension,'' \emph{J. of the Amer. Math.
  Soc.}, vol.~22, no.~1, pp. 1--53, 2009.

\bibitem{Remi2011}
R.~Gribonval, ``Should penalized least squares regression be interpreted as
  maximum a posteriori estimation?'' \emph{IEEE Trans. on Signal Process.},
  vol.~59, no.~5, pp. 2405--2410, May 2011.

\bibitem{Donoho2010}
A.~Maleki and D.~Donoho, ``Optimally tuned iterative reconstruction algorithms
  for compressed sensing,'' \emph{IEEE J. of Select. Topics in Signal
  Process.}, vol.~4, no.~2, pp. 330--341, April 2010.

\bibitem{Donoho_MAI2012}
D.~Donoho, Y.~Tsaig, I.~Drori, and J.-L. Starck, ``Sparse solution of
  underdetermined systems of linear equations by stagewise orthogonal matching
  pursuit,'' \emph{IEEE Trans. on Inform. Theory}, vol.~58, no.~2, pp.
  1094--1121, Feb 2012.

\bibitem{Norbert2014}
N.~Goertz, C.~Guo, A.~Jung, M.~Davies, and G.~Doblinger, ``Iterative recovery
  of dense signals from incomplete measurements,'' \emph{IEEE Signal Process.
  Lett.}, vol.~21, no.~9, pp. 1059--1063, Sept 2014.

\bibitem{gribonval_compressible11}
R.~Gribonval, V.~Cehver, and M.~Davies, ``Compressible distributions for high
  dimensional statistics,'' \emph{IEEE Trans. on Inform. Theory}, vol.~58,
  no.~8, pp. 5016--5034, Aug 2012.

\bibitem{BergFriedlander2008}
\BIBentryALTinterwordspacing
E.~van~den Berg and M.~P. Friedlander, ``Probing the pareto frontier for basis
  pursuit solutions,'' \emph{SIAM J. on Scientific Computing}, vol.~31, no.~2,
  pp. 890--912, 2008. [Online]. Available:
  \url{http://link.aip.org/link/?SCE/31/890}
\BIBentrySTDinterwordspacing

\end{thebibliography}

\end{document}